\documentclass[journal]{IEEEtran}
\usepackage[dvips]{graphicx}
\usepackage{epsfig,multirow}
\usepackage{amsmath,amssymb,bm}
\usepackage{amsfonts,dsfont,color,bbm}
\usepackage{algorithm}
\usepackage[noend]{algpseudocode}
\usepackage[font=footnotesize,labelfont=bf,justification=justified,singlelinecheck=false]{caption}
\usepackage{subcaption,slashbox}
\usepackage{cite,epstopdf,epsf,epsfig}
\usepackage{url}
\usepackage{setspace}

\newcommand{\subparagraph}{}
\renewcommand{\vec}[1]{\mbox{\boldmath${#1}$}}

\newcommand{\be}{\begin{equation}}
\newcommand{\ee}{\end{equation}}
\newcommand{\bea}{\begin{eqnarray}}
\newcommand{\eea}{\end{eqnarray}}

\newcommand{\ei}{\end{itemize}}
\newcommand{\bi}{\begin{itemize}}

\let\oldReturn\Return
\renewcommand{\Return}{\State\oldReturn}

\DeclareMathSizes{10}{9}{6}{4}
\AtBeginDocument{
\addtolength{\abovedisplayskip}{-0.9ex}
\addtolength{\abovedisplayshortskip}{-0.4ex}
\addtolength{\belowdisplayskip}{-0.9ex}
\addtolength{\belowdisplayshortskip}{-0.4ex}
\addtolength{\belowcaptionskip}{-4.9ex}
}

\begin{document}

\title{Online Training of LSTM Networks in Distributed Systems for Variable Length Data Sequences}

\author{Tolga Ergen and Suleyman S. Kozat \textit{Senior Member, IEEE}\thanks{This work is supported in part by TUBITAK Contract No 115E917.}
\thanks{The authors are with the Department of Electrical and Electronics Engineering, Bilkent University, Bilkent, Ankara 06800, Turkey, Tel: +90 (312) 290-2336, Fax: +90 (312) 290-1223, {(contact e-mail: \{ergen, kozat\}@ee.bilkent.edu.tr)}.} }
\maketitle
\begin{abstract}
In this brief paper, we investigate online training of Long Short Term Memory (LSTM) architectures in a distributed network of nodes, where each node employs an LSTM based structure for online regression. In particular, each node sequentially receives a variable length data sequence with its label and can only exchange information with its neighbors to train the LSTM architecture. We first provide a generic LSTM based regression structure for each node. In order to train this structure, we put the LSTM equations in a nonlinear state space form for each node and then introduce a highly effective and efficient Distributed Particle Filtering (DPF) based training algorithm. We also introduce a Distributed Extended Kalman Filtering (DEKF) based training algorithm for comparison. Here, our DPF based training algorithm guarantees convergence to the performance of the optimal LSTM coefficients in the mean square error (MSE) sense under certain conditions. We achieve this performance with communication and computational complexity in the order of the first order gradient based methods. Through both simulated and real life examples, we illustrate significant performance improvements with respect to the state of the art methods. 
\end{abstract}
\begin{keywords}
Distributed learning, online learning, particle filtering, extended Kalman filtering, LSTM networks.
\end{keywords}
\section{Introduction}
Neural networks provide enhanced performance for a wide range of engineering applications, e.g., prediction \cite{neural1} and human behavior modeling \cite{human}, thanks to their highly strong nonlinear modeling capabilities. Among neural networks, especially recurrent neural networks (RNNs) are used to model time series and temporal data due to their inherent memory storing the past information \cite{tsoi}. However, since simple RNNs lack control structures, the norm of gradient may grow or decay in a fast manner during training, i.e., the exploding and vanishing gradient issues \cite{rnndisadvantage}. Due to these problems, simple RNNs are insufficient to capture long and short term dependencies \cite{rnndisadvantage}. To circumvent this issue, a novel RNN architecture with control structures, i.e., the Long Short Term Memory (LSTM) network, is introduced \cite{hoch_lstm}. However, since LSTM networks have additional nonlinear control structures with several parameters, they may also suffer from training issues \cite{hoch_lstm}. 

To this end, in this brief paper, we consider online training of the parameters of an LSTM structure in a distributed network of nodes. Here, we have a network of nodes, where each node has a set of neighboring nodes and can only exchange information with these neighbors. In particular, each node sequentially receives a variable length data sequence with its label and trains the parameters of the LSTM network. Each node can also communicate with its neighbors to share information in order to enhance the training performance since the goal is to train one set of LSTM coefficients using all the available data. As an example application, suppose that we have a database of labelled tweets and our aim is to train an emotion recognition engine based on an LSTM structure, where the training is performed in an online and distributed manner using several processing units. Words in each tweet are represented by word2vec vectors \cite{tweet} and tweets are distributed to several processing units in an online manner.

The LSTM architectures are usually trained in a batch setting in the literature, where all data instances are present and processed together \cite{tsoi}. However, for applications involving big data, storage issues may arise due to keeping all the data in one place \cite{batch}. Additionally, in certain frameworks, all data instances are not available beforehand since instances are received in a sequential manner, which precludes batch training \cite{batch}. Hence, we consider online training, where we sequentially receive the data to train the LSTM architecture without storing the previous data instances. Note that even though we work in an online setting, we may still suffer from computational power and storage issues due to large amount of data \cite{bigdata}. As an example, in tweet emotion recognition applications, the systems are usually trained using an enormous amount of data to achieve sufficient performance, especially for agglutinative languages \cite{tweet}. For such tasks distributed architectures are used. In this basic distributed architectures, commonly named as centralized approach \cite{bigdata},  the whole data is distributed to different nodes and trained parameters are merged later at a central node \cite{tsoi}. However, this centralized approach requires high storage capacity and computational power at the central node \cite{bigdata}. Additionally, centralized strategies have a potential risk of failure at the central node. To circumvent these issues, we distribute both the processing as well as the data to all the nodes and allow communication only between neighboring nodes, hence, we remove the need for a central node. In particular, each node sequentially receives a variable length data sequence with its label and exchanges information only with its neighboring nodes to train the common LSTM parameters.

For online training of the LSTM architecture in a distributed manner, one can employ one of the first order gradient based algorithms at each node due to their efficiency \cite{tsoi} and exchange estimates among neighboring nodes as in \cite{distributed_gd_convergence}. However, since these training methods only exploit the first order gradient information, they suffer from poor performance and convergence issues. As an example, the Stochastic Gradient Descent (SGD) based algorithms usually have slower convergence compared to the second order methods \cite{distributed_gd_convergence,tsoi}. On the other hand, the second order gradient based methods require much higher computational complexity and communication load while providing superior performance compared to the first order methods \cite{tsoi}. Following the distributed implementation of the first order methods, one can implement the second order training methods in a distributed manner, where we share not only the estimates but also the Jacobian matrix, e.g.,  the Distributed Extended Kalman Filtering (DEKF) algorithm \cite{distributed_kalman,optimal}. However, as in the first order case, these sharing and combining the information at each node is adhoc, which does not provide the optimal training performance \cite{distributed_kalman}. 
In this brief paper, to provide improved performance with respect to the second order methods while preserving both communication and computational complexity similar to the first order methods, we introduce a highly effective distributed online training method based on the particle filtering algorithm \cite{markovconvergence}. We first propose an LSTM based model for variable length data regression. We then put this model in a nonlinear state space form to train the model in an online and optimal manner. 

Our main contributions include: 1) We introduce distributed LSTM training methods in an online setting for variable length data sequences. Our Distributed Particle Filtering (DPF) based training algorithm guarantees convergence to the optimal centralized training performance in the mean square error (MSE) sense; 2) We achieve this performance with a computational complexity and a communication load in the order of the first order gradient based methods; 3) Through simulations involving real life and financial data, we illustrate significant performance improvements with respect to the state of the art methods \cite{ekf_lstm2,rtrl}.

The organization of this brief paper is as follows. In Section \ref{sec:problemdescription}, we first describe the variable length data regression problem in a network of nodes and then introduce an LSTM based structure. Then, in Section \ref{sec:main}, we first put this structure in a nonlinear state space form and then introduce our training algorithms. In Section \ref{sec:simulations}, we illustrate the merits of our algorithms through simulations. We then finalize the brief paper with concluding remarks in Section \ref{sec:conclusion}.
\section{Model and Problem Description}\label{sec:problemdescription}
Here\footnote{All column vectors (or matrices) are denoted by boldface lower (or uppercase) case letters. For a matrix  $\vec{A}$ (or a vector $\vec{a}$), $\vec{A}^T$ ($\vec{a}^T$) is its ordinary transpose. The time index is given as subscript, e.g., $\vec{u}_t$ is the vector at time $t$. Here, $\vec{1}$ (or $\vec{0}$) is a vector of all ones (or zeros) and $\vec{I}$ is the identity matrix, where the sizes of these notations are understood from the context.}, we consider a network of $K$ nodes. In this network, we declare two nodes that can exchange information as neighbors and denote the neighborhood of each node $k$ as $\mathcal{N}_k$ that also includes the node $k$, i.e., $k \in \mathcal{N}_k$. At each node $k$, we sequentially receive $\{d_{k,t}\}_{t\geq 1}$, $d_{k,t} \in \mathbb{R}$ and matrices, $\{\vec{X}_{k,t} \}_{t \geq 1}$, defined as $\vec{X}_{k,t}=[\vec{x}_{k,t}^{(1)} \text{ } \vec{x}_{k,t}^{(2)}\ldots \vec{x}_{k,t}^{(m_t)} ]$,
where $\vec{x}_{k,t}^{(l)} \in \mathbb{R}^{p}$, $\forall l \in \lbrace1,2, \ldots, m_t\rbrace$ and $m_{t} \in \mathbb{Z}^{+}$ is the number of columns in $\vec{X}_{k,t}$, which can change with respect to $t$. In our network, each node $k$ aims to learn a certain relation between the desired value $d_{k,t}$ and matrix $\vec{X}_{k,t}$. After observing $\vec{X}_{k,t}$ and $d_{k,t}$, each node $k$ first updates its belief about the relation and then exchanges an updated information with its neighbors. After receiving $\vec{X}_{k,t+1}$, each node $k$ estimates the next signal $d_{k,t+1}$ as $\hat{d}_{k,t+1}$.
Based on $d_{k,t+1}$, each node $k$ suffers the loss $l(d_{k,t+1}, \hat{d}_{k,t+1})$ at time instance $t+1$. This framework models a wide range of applications in the machine learning and signal processing literatures, e.g., sentiment analysis \cite{tweet}. As an example, in tweet emotion recognition application \cite{tweet}, each $\vec{X}_{k,t}$ corresponds to a tweet, i.e., the $t$\textsuperscript{th} tweet at the node (processing unit) $k$. For the $t$\textsuperscript{th} tweet at the node $k$, one can construct $\vec{X}_{k,t}$ by finding word2vec representation of each word, i.e., $\vec{x}_{k,t}^{(l)}$ for the $l$\textsuperscript{th} word. After receiving $d_{k,t}$, i.e., the desired emotion label for the $t$\textsuperscript{th} tweet at the node $k$, each node $k$ first updates its belief about the relation between the tweet and its emotion label, and then exchanges information, e.g., the trained system parameters, with its neighboring units to estimate the next label. 
\begin{figure}[t]
 \centering 
 \includegraphics[width=0.40\textwidth, height=0.25\textwidth]{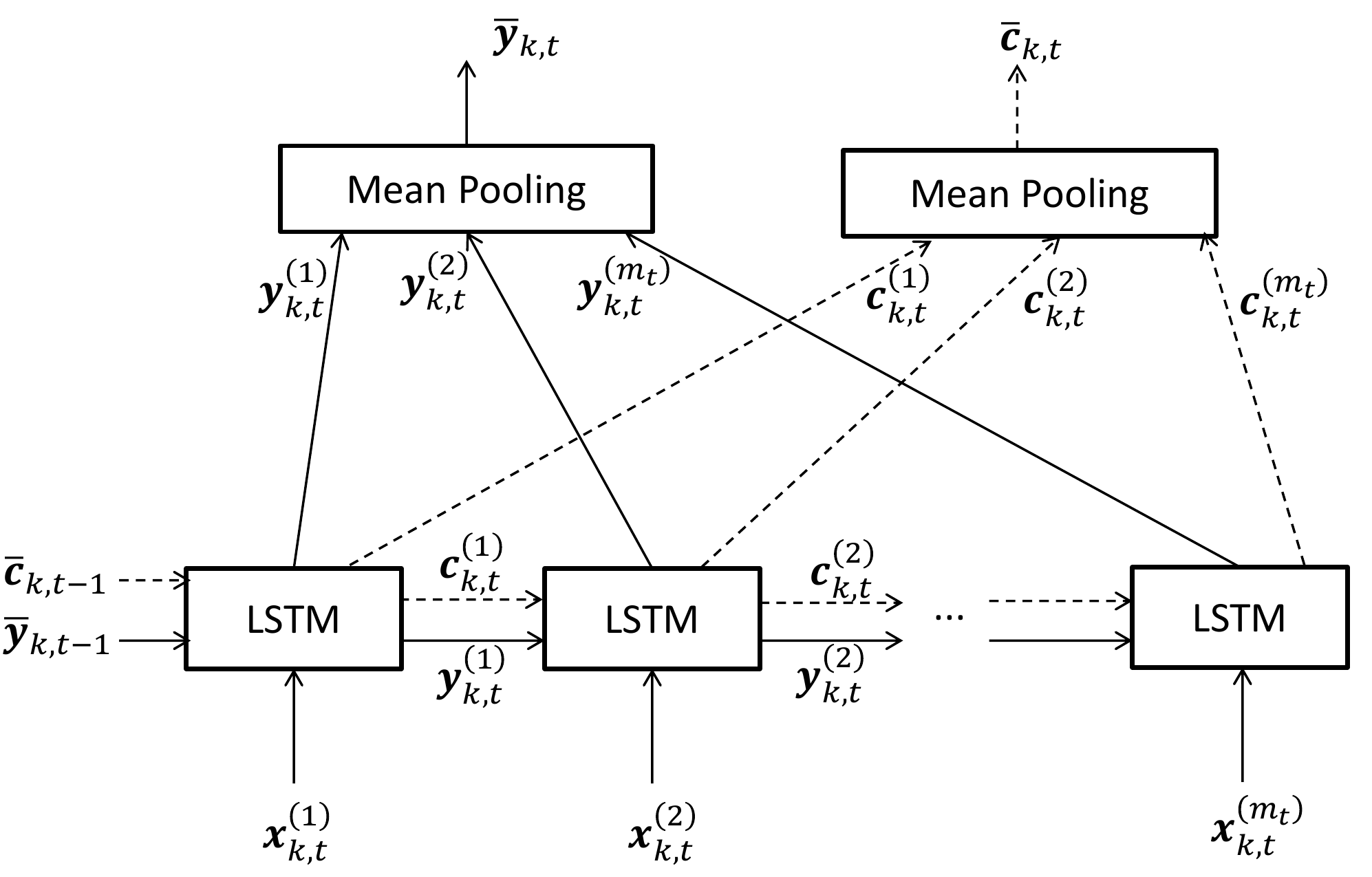}\\
 \caption{Detailed schematic of each node $k$ in our network.\\\\ } \label{meanpooling}
\end{figure}

In this brief paper, each node $k$ generates an estimate $\hat{d}_{k,t}$ using the LSTM architecture. Although there exist different variants of LSTM, we use the most widely used variant \cite{hoch_lstm}, i.e., the LSTM architecture without peephole connections. The input $\vec{X}_{k,t}$ is first fed to the LSTM architecture as illustrated in Fig. \ref{meanpooling}, where the internal equations are given as \cite{hoch_lstm}:
\begin{align}
  & \vec{i}_{k,t}^{(l)} = \sigma(\vec{W}_{k}^{(i)}\vec{x}_{k,t}^{(l)}+\vec{R}_{k}^{(i)}\vec{y}_{k,t}^{(l-1)}+\vec{b}_{k}^{(i)}) \label{eq:i} \\
    & \vec{f}_{k,t}^{(l)} = \sigma(\vec{W}_{k}^{(f)}\vec{x}_{k,t}^{(l)}+\vec{R}_{k}^{(f)}\vec{y}_{k,t}^{(l-1)}+\vec{b}_{k}^{(f)}) \label{eq:f} \\
  & \vec{c}_{k,t}^{(l)} = \vec{i}_{k,t}^{(l)} \odot g(\vec{W}_{k}^{(z)}\vec{x}_{k,t}^{(l)}+\vec{R}_{k}^{(z)}\vec{y}_{k,t}^{(l-1)}+\vec{b}_{k}^{(z)})+ \vec{f}_{k,t}^{(l)} \odot \vec{c}_{k,t}^{(l-1)} \label{eq:state} \\
  & \vec{o}_{k,t}^{(l)} = \sigma(\vec{W}_{k}^{(o)}\vec{x}_{k,t}^{(l)}+\vec{R}_{k}^{(o)}\vec{y}_{k,t}^{(l-1)}+ \vec{b}_{k}^{(o)}) \label{eq:o} \\
  & \vec{y}_{k,t}^{(l)} = \vec{o}_{k,t}^{(l)} \odot h(\vec{c}_{k,t}^{(l)}), \label{eq:output}
\end{align}
where $\vec{x}_{k,t}^{(l)} \in \mathbb{R}^p$ is the input vector, $\vec{y}_{k,t}^{(l)} \in \mathbb{R}^n$ is the output vector and $\vec{c}_{k,t}^{(l)}\in \mathbb{R}^n$ is the state vector for the $l$\textsuperscript{th} LSTM unit. Moreover, $\vec{o}_{k,t}^{(l)}$, $\vec{f}_{k,t}^{(l)}$  and $\vec{i}_{k,t}^{(l)}$ represent the output, forget and input gates, respectively. $g(\cdot)$ and $h(\cdot)$ are set to the hyperbolic tangent function and apply vectors pointwise. Likewise, $\sigma(\cdot)$ is the pointwise sigmoid function. The operation $\odot$ represents the elementwise multiplication
of two vectors of the same size. As the coefficient matrices and the weight vectors of the LSTM architecture, we have $\vec{W}_{k}^{(.)}$, $\vec{R}_{k}^{(.)}$ and $\vec{b}_{k}^{(.)} $, where the sizes are chosen according to the input and output vectors. Given the outputs of LSTM for each column of $\vec{X}_{k,t}$ as seen in Fig. \ref{meanpooling}, we generate the estimate for each node $k$ as follows
\begin{align}
\hat{d}_{k,t}=\vec{w}_{k,t}^{T}\vec{\bar{y}}_{k,t}, \label{out}
\end{align}
where $\vec{w}_{k,t} \in \mathbb{R}^n$ is a vector of the regression coefficients and $\vec{\bar{y}}_{k,t} \in \mathbb{R}^n$ is a vector obtained by taking average of the LSTM outputs for each column of $\vec{X}_{k,t}$, i.e., known as the mean pooling method, as described in Fig. \ref{meanpooling}. 

\noindent
{\bf Remark 1:} In \eqref{out}, we use the mean pooling method to generate $\vec{\bar{y}}_{k,t}$. One can also use the other pooling methods by changing the calculation of $\vec{\bar{y}}_{k,t}$ and then generate the estimate as in \eqref{out}. As an example, for the max and last pooling methods, we use $\vec{\bar{y}}_{k,t}=\max_{i} \vec{y}_{k,t}^{(i)}$ and $\vec{\bar{y}}_{k,t}=\vec{y}_{k,t}^{(m_t)}$, respectively. All our derivations hold for these pooling methods and the other LSTM architectures. We provide the required updates for different LSTM architectures in the next section.
\section{Online Distributed Training Algorithms}\label{sec:main}
In this section, we first give the LSTM equations for each node in a nonlinear state space form. Based on this form, we then introduce our distributed algorithms to train the LSTM parameters in an online manner.

Considering our model in Fig. \ref{meanpooling} and  the LSTM equations in \eqref{eq:i}, \eqref{eq:f}, \eqref{eq:state}, \eqref{eq:o} and \eqref{eq:output}, we have the following nonlinear state space form for each node $k$
\begin{align}
\label{state1}
&\vec{\bar{c}}_{k,t}=\Omega(\vec{\bar{c}}_{k,t-1},\vec{X}_{k,t}, \vec{\bar{y}}_{k,t-1})\\
\label{state2} 
&\vec{\bar{y}}_{k,t}=\Theta(\vec{\bar{c}}_{k,t},\vec{X}_{k,t}, 
 \vec{\bar{y}}_{k,t-1})\\ \label{state3} 
&\vec{\theta}_{k,t}=\vec{\theta}_{k,t-1}\\\label{output}
&d_{k,t}=\vec{w}_{k,t}^{T}\vec{\bar{y}}_{k,t}+\varepsilon_{k,t},
\end{align}
where $\Omega(\cdot)$ and $\Theta(\cdot)$ represent the nonlinear mappings performed by the consecutive LSTM units and the mean pooling operation as illustrated in Fig. \ref{meanpooling}, and $\vec{\theta}_{k,t} \in \mathbb{R}^{n_\theta}$ is a parameter vector consisting of{\def\OldComma{,}
    \catcode`\,=13
    \def,{%
      \ifmmode%
        \OldComma\discretionary{}{}{}%
      \else%
        \OldComma%
      \fi    }%
    $\lbrace\vec{w}_{k}, \vec{W}_{k}^{(z)}, \vec{R}_{k}^{(z)}, \vec{b}_{k}^{(z)}, \vec{W}_{k}^{(i)}, \vec{R}_{k}^{(i)}, \vec{b}_{k}^{(i)}, \vec{W}_{k}^{(f)}, \vec{R}_{k}^{(f)}, \vec{b}_{k}^{(f)}, \vec{W}_{k}^{(o)}, \vec{R}_{k}^{(o)}, \vec{b}_{k}^{(o)} \rbrace$}, where $n_{\theta}=4n(n+p)+5n$. Since the LSTM parameters are the states of the network to be estimated, we also include the static equation \eqref{state3} as our state. Furthermore, $\varepsilon_{k,t}$ represents the error in observations and it is a zero mean Gaussian random variable with variance $R_{k,t}$.
    
\noindent
{\bf Remark 2:} We can also apply the introduced algorithms to different implementations of the LSTM architecture \cite{hoch_lstm}. For this purpose, we modify the function $\Omega(\cdot)$ and $\Theta(\cdot)$ in \eqref{state1} and \eqref{state2} according to the chosen LSTM architecture. We also alter  $\vec{\theta}_{k,t}$ in \eqref{state3} by adding or removing certain parameters according to the chosen LSTM architecture. 
\subsection{Online Training Using the DEKF Algorithm}
In this subsection, we first derive our training method based on the EKF algorithm, where each node trains its LSTM parameters without any communication with its neighbors. We then introduce our training method based on the DEKF algorithm in order to train the LSTM architecture when we allow communication between the neighbors.

The EKF algorithm is based on the assumption that the state distribution given the observations is Gaussian \cite{optimal}. To meet this assumption, we introduce Gaussian noise to \eqref{state1}, \eqref{state2} and \eqref{state3}. By this, we have the following model for each node $k$
\begin{align}
\label{ekf_compact1}
&\begin{bmatrix}\vec{\bar{c}}_{k,t}\\
\vec{\bar{y}}_{k,t}\\
\vec{\theta}_{k,t}
\end{bmatrix}=
\begin{bmatrix}
\Omega(\vec{\bar{c}}_{k,t-1},\vec{X}_{k,t}, \vec{\bar{y}}_{k,t-1})\\ 
\Theta(\vec{\bar{c}}_{k,t},\vec{X}_{k,t}, \vec{\bar{y}}_{k,t-1})\\
\vec{\theta}_{k,t-1}
\end{bmatrix}+
\begin{bmatrix}
\vec{e}_{k,t}\\
\vec{\epsilon}_{k,t}\\
\vec{\upsilon}_{k,t}
\end{bmatrix}\\ \label{ekf_compact2}
&d_{k,t}=\vec{w}_{k,t}^{T}\vec{\bar{y}}_{k,t}+\varepsilon_{k,t},
\end{align}
where $[\vec{e}_{k,t}^{T}, \vec{\epsilon}_{k,t}^{T}, \vec{\upsilon}_{k,t}^{T}]^{T}$ is zero mean Gaussian process with covariance $\vec{Q}_{k,t}$. Here, each node $k$ is able to observe only $d_{k,t}$ to estimate $\vec{\bar{c}}_{k,t}$, $\vec{\bar{y}}_{k,t}$ and $\vec{\theta}_{k,t}$. Hence, we group $\vec{\bar{c}}_{k,t}$, $\vec{\bar{y}}_{k,t}$ and $\vec{\theta}_{k,t}$ together into a vector as the hidden states to be estimated.
\subsubsection{Online Training with the EKF Algorithm:\text{}} In this subsection, we derive the online training method based on the EKF algorithm when we do not allow communication between the neighbors. Since the system in \eqref{ekf_compact1} and \eqref{ekf_compact2} is already in a nonlinear state space form, we can directly apply the EKF algorithm \cite{optimal} as follows
\begin{align}
&\text{\underline{Time Update}: } \nonumber  \\
 \label{ekf4}
 &\vec{\bar{c}}_{k,t|t-1}=\Omega(\vec{\bar{c}}_{k,t-1|t-1},\vec{X}_{k,t}, \vec{\bar{y}}_{k,t-1|t-1}) \\
 \label{ekf5}
 &\vec{\bar{y}}_{k,t|t-1}=\Theta(\vec{\bar{c}}_{t|t-1},\vec{X}_{k,t}, \vec{\bar{y}}_{k,t-1|t-1}) \\
 \label{theta}
&\vec{\theta}_{k,t|t-1}=\vec{\theta}_{k,t-1|t-1}\end{align}\begin{align}
   \label{ekf6}
&\vec{\Sigma}_{k,t|t-1}=\vec{F}_{k,t-1}\vec{\Sigma}_{k,t-1|t-1}\vec{F}_{k,t-1}^{T}+\vec{Q}_{k,t-1}
\\&\text{\underline{Measurement Update}: } \nonumber \\
& R=\vec{H}_{k,t}^{T}\vec{\Sigma}_{k,t|t-1}\vec{H}_{k,t}+R_{k,t} \nonumber \\
&\begin{bmatrix}\vec{\bar{c}}_{k,t|t}\\
\vec{\bar{y}}_{k,t|t}\\
\vec{\theta}_{k,t|t}
\end{bmatrix}=\begin{bmatrix}\vec{\bar{c}}_{k,t|t-1} \nonumber\\
\vec{\bar{y}}_{k,t|t-1}\\
\vec{\theta}_{k,t|t-1}
\end{bmatrix}+\vec{\Sigma}_{k,t|t-1}\vec{H}_{k,t} R^{-1}(d_{k,t}-\hat{d}_{k,t}) \nonumber\\
&\vec{\Sigma}_{k,t|t}=\vec{\Sigma}_{k,t|t-1}-\vec{\Sigma}_{k,t|t-1}\vec{H}_{k,t} R^{-1}\vec{H}_{k,t}^{T}\vec{\Sigma}_{k,t|t-1}, \nonumber
\end{align}
where $\vec{\Sigma}\in\mathbb{R}^{(2n+n_{\theta})\times(2n+n_{\theta})}$ is the error covariance matrix, $\vec{Q}_{k,t} \in \mathbb{R}^{(2n+n_{\theta})\times (2n+n_{\theta})}$ is the state noise covariance and $R_{k,t} \in \mathbb{R}$  is the measurement noise variance. Additionally, we assume that $R_{k,t}$ and $\vec{Q}_{k,t}$ are known terms. We compute $\vec{H}_{k,t}$ and $\vec{F}_{k,t}$ as follows
\begin{align}
\label{ekf_h}\vec{H}_{k,t}^{T}=\begin{bmatrix}
 \frac{\partial  \hat{d}_{k,t}}{\partial \vec{\bar{c}}} && \frac{\partial \hat{d}_{k,t}}{\partial \vec{\bar{y}}}  && \frac{\partial  \hat{d}_{k,t}}{\partial \vec{\theta}}
\end{bmatrix}\Bigr|_{\substack{\vec{\bar{c}}=\vec{\bar{c}}_{k,t|t-1} \\ \vec{\bar{y}}=\vec{\bar{y}}_{k,t|t-1} \\ \vec{\theta}=\vec{\theta}_{k,t|t-1}}}
\end{align}
and
\begin{align}
\label{ekf_f}\vec{F}_{k,t}=\begin{bmatrix}
\frac{\partial \Omega(\vec{\bar{c}},\vec{X}_{k,t}, \vec{\bar{y}})}{\partial \vec{\bar{c}}} &\frac{\partial \Omega(\vec{\bar{c}},\vec{X}_{k,t}, \vec{\bar{y}})}{\partial \vec{\bar{y}}}  & \frac{\partial \Omega(\vec{\bar{c}},\vec{X}_{k,t}, \vec{\bar{y}})}{\partial \vec{\theta}} \\ 
\frac{\partial \Theta(\vec{\bar{c}},\vec{X}_{k,t}, \vec{\bar{y}})}{\partial \vec{\bar{c}}} &\frac{\partial \Theta(\vec{\bar{c}},\vec{X}_{k,t}, \vec{\bar{y}})}{\partial \vec{\bar{y}}}  & \frac{\partial \Theta(\vec{\bar{c}},\vec{X}_{k,t}, \vec{\bar{y}})}{\partial \vec{\theta}} \\
\vec{0} & \vec{0} & \vec{I}
\end{bmatrix}\Biggr|_{\substack{\vec{\bar{c}}=\vec{\bar{c}}_{k,t|t}\\ \vec{\bar{y}}=\vec{\bar{y}}_{k,t|t}\\ \vec{\theta}=\vec{\theta}_{k,t|t}}},
\end{align}
where $\vec{F}_{k,t} \in \mathbb{R}^{(2n+n_{\theta})\times (2n+n_{\theta})}$ and $\vec{H}_{k,t} \in \mathbb{R}^{ (2n+n_{\theta})}$. 
\subsubsection{Online Training with the DEKF Algorithm:\text{}} In this subsection, we introduce our online training method based on the DEKF algorithm for the network described by \eqref{ekf_compact1} and \eqref{ekf_compact2}. In our network of $K$ nodes, we denote the number of neighbors for the node $k$ as $\eta_k$, i.e., also called as the degree of the node $k$ \cite{distributed_kalman}. With this structure, the time update equations in \eqref{ekf4}, \eqref{ekf5}, \eqref{theta} and \eqref{ekf6} still hold for each node $k$. However, since we have information exchange between the neighbors, the measurement update equations of each node $k$ adopt the iterative scheme \cite{distributed_kalman} as the following.
\begin{align*}
&\text{For the node }k\text{ at time } t \text{:}   \nonumber \\
&\vec{\phi}_{k,t}\longleftarrow [\vec{\bar{c}}_{k,t|t-1}^T \text{ }
\vec{\bar{y}}_{k,t|t-1}^T \text{ }\vec{\theta}_{k,t|t-1}^T ]^T\\
&\vec{\Phi}_{k,t}\longleftarrow\vec{\Sigma}_{k,t|t-1} \nonumber\\
&\text{For each } l\in \mathcal{N}_k \text{ repeat: }\nonumber\\
&\hspace*{20pt}  R\longleftarrow\vec{H}_{l,t}^{T}\vec{\Phi}_{k,t}\vec{H}_{l,t}+R_{l,t} \\
&\hspace*{20pt} \vec{\phi}_{k,t}\longleftarrow\vec{\phi}_{k,t}+\vec{\Phi}_{k,t}\vec{H}_{l,t} R^{-1}(d_{l,t}-\vec{w}_{k,t|t-1}^{T}\vec{\bar{y}}_{k,t|t-1}) \\
&\hspace*{20pt} \vec{\Phi}_{k,t}\longleftarrow\vec{\Phi}_{k,t}-\vec{\Phi}_{k,t}\vec{H}_{l,t} R^{-1}\vec{H}_{l,t}^{T}\vec{\Phi}_{k,t}.
\end{align*}
Now, we update the state and covariance matrix estimate as
\begin{align*}
&\vec{\Sigma}_{k,t|t}=\vec{\Phi}_{k,t}\\
& [\vec{\bar{c}}_{k,t|t}^T \text{ }
\vec{\bar{y}}_{k,t|t}^T \text{ }
\vec{\theta}_{k,t|t} ^T]^T= \sum _{l \in \mathcal{N}_{k}} c(k,l) \vec{\phi}_{l,t},
\end{align*}
where $c(k,l)$ is the weight between the node $k$ and $l$ and we compute these weights using the Metropolis rule as follows
\begin{equation} \label{eq:phi}
c(k,l) =
\begin{cases} 
1/\max(\eta_k,\eta_l)\hspace{.3cm} & \text{ if } l \in \mathcal{N}_k/k\\     1- \sum_{l \in \mathcal{N}_k/k}c(k,l) & \text{ if } k=l  \\ 0\hspace{1.5cm}& \text{ if }  l \notin \mathcal{N}_k \
\end{cases}.
\end{equation}
With these steps, we can update all the nodes in our network as illustrated in Algorithm \ref{alg1}.

According to the procedure in Algorithm \ref{alg1}, the computational complexity of our training method results in $\mathcal{O}(\eta_k( n^8+n^4 p^4))$ computations at each node $k$ due to matrix and vector multiplications on lines \ref{comp1} and \ref{comp2} as shown in Table \ref{tab:complexity}.
\begin{algorithm}
\footnotesize
\caption{Training based on the DEKF Algorithm}
\label{alg1}
\begin{algorithmic}[1]
\State{According to \eqref{ekf_h}, compute $\vec{H}_{k,t}$, $\forall k \in \lbrace
 1,2, \ldots, K \rbrace$}
\For{$k=1:K$}
\State {$
\vec{\phi}_{k,t}\longleftarrow [\vec{\bar{c}}_{k,t|t-1}^T \text{ }\vec{\bar{y}}_{k,t|t-1}^T  \text{ }\vec{\theta}_{k,t|t-1}^T ]^T$}
\State{$\vec{\Phi}_{k,t}\longleftarrow\vec{\Sigma}_{k,t|t-1} $}
    \For{$l\in \mathcal{N}_k$}
       \State{$R\longleftarrow\vec{H}_{l,t}^{T}\vec{\Phi}_{k,t}\vec{H}_{l,t}+R_{l,t}$}
       \State{$\vec{\phi}_{k,t}\longleftarrow\vec{\phi}_{k,t}+\vec{\Phi}_{k,t}\vec{H}_{l,t} R^{-1}(d_{l,t}-\vec{w}_{k,t|t-1}^{T}\vec{\bar{y}}_{k,t|t-1}) $} \label{ekf_comp1}
       \State{$\vec{\Phi}_{k,t}\longleftarrow\vec{\Phi}_{k,t}-\vec{\Phi}_{k,t}\vec{H}_{l,t} R^{-1}\vec{H}_{l,t}^{T}\vec{\Phi}_{k,t}$} \label{comp1}
    \EndFor
 \State {\textbf{end for}}
\EndFor
 \State {\textbf{end for}}
 \For{$k=1:K$}
 \State{Using \eqref{eq:phi}, calculate $c(k,l) \text{, }\forall l\in \mathcal{N}_k$}
\State $ [\vec{\bar{c}}_{k,t|t}^T \text{ }\vec{\bar{y}}_{k,t|t}^T  \text{ }\vec{\theta}_{k,t|t}^T ]^T \longleftarrow \sum _{l \in \mathcal{N}_{k}} c(k,l) \vec{\phi}_{l,t}$ \label{ekf_comp2}
\State{$\vec{\Sigma}_{k,t|t} \longleftarrow\vec{\Phi}_{k,t}$}
\State{According to \eqref{ekf_f}, compute $\vec{F}_{k,t}$}
\State $\vec{\bar{c}}_{k,t+1|t}\longleftarrow\Omega(\vec{\bar{c}}_{k,t|t},\vec{X}_{k,t}, \vec{\bar{y}}_{k,t|t}) $
\State$\vec{\bar{y}}_{k,t+1|t}\longleftarrow\Theta(\vec{\bar{c}}_{k,t+1|t},\vec{X}_{k,t}, \vec{\bar{y}}_{k,t|t}) $
\State $\vec{\theta}_{k,t+1|t}\longleftarrow\vec{\theta}_{k,t|t} $
\State$\vec{\Sigma}_{k,t+1|t}\longleftarrow\vec{F}_{k,t}\vec{\Sigma}_{k,t|t}\vec{F}_{k,t}^{T}+\vec{Q}_{k,t}$ \label{comp2}
\EndFor
\State {\textbf{end for}}
\end{algorithmic}
\end{algorithm}
\subsection{Online Training Using the DPF Algorithm}
In this subsection, we first derive our training method based on the PF algorithm when we do not allow communication between the nodes. We then introduce our online training method based on the DPF algorithm when the nodes share information with their neighbors. 

The PF algorithm only requires the independence of the noise samples in \eqref{ekf_compact1} and \eqref{ekf_compact2}. Thus, we modify our system in \eqref{ekf_compact1} and \eqref{ekf_compact2} for the node $k$ as follows
\begin{align}
&\vec{a}_{k,t}=\varphi(\vec{a}_{k,t-1},\vec{X}_{k,t})+\vec{\gamma}_{k,t} \label{pfcompact1} \\
&d_{k,t}=\vec{w}_{k,t}^{T}\vec{\bar{y}}_{k,t}+\varepsilon_{k,t}, \label{pfcompact2}
\end{align}
where $\vec{\gamma}_{k,t}$ and $\varepsilon_{k,t}$ are independent state and measurement noise samples, respectively, $\varphi(\cdot, \cdot)$ is the nonlinear mapping in \eqref{ekf_compact1} and $\vec{a}_{k,t}\triangleq[\vec{\bar{c}}_{k,t}^T \text{ }\vec{\bar{y}}_{k,t}^T \text{ }\vec{\theta}_{k,t}^T ]^T$.
\subsubsection{Online Training with the PF Algorithm:\text{ }}
For the system in \eqref{pfcompact1} and \eqref{pfcompact2}, our aim is to obtain $\mathbf{E}[\vec{a}_{k,t}| d_{k,1:t}]$, i.e., the optimal estimate for the hidden state in the MSE sense. To achieve this, we first obtain posterior distribution of the states, i.e., $p(\vec{a}_{k,t} | d_{k,1:t})$. Based on the posterior density function, we then calculate the conditional mean estimate. In order to obtain the posterior distribution, we apply the PF algorithm \cite{djuric}.

In this algorithm, we have the samples and the corresponding weights of $p(\vec{a}_{k,t} | d_{k,1:t})$, i.e., denoted as $\lbrace \vec{a}_{k,t}^{i}, \omega_{k,t}^{i} \rbrace_{i=1}^{N}$. Based on the samples, we obtain the posterior distribution as follows
\begin{align}
p(\vec{a}_{k,t} | d_{k,1:t})\approx \sum_{i=1}^{N} \omega_{k,t}^{i}\delta(\vec{a}_{k,t}-\vec{a}_{k,t}^{i}). \label{distribution}
\end{align}
Sampling from the desired distribution $p(\vec{a}_{k,t}|d_{k,1:t})$ is intractable in general so that we obtain the samples from $q(\vec{a}_{k,t}| d_{k,1:t})$, which is called as importance function \cite{djuric}. To calculate the weights in \eqref{distribution}, we use the following formula
\begin{align}
w_{k,t}^{i} \propto \frac{p(\vec{a}_{k,t}^{i}| d_{k,1:t})}{q(\vec{a}_{k,t}^{i}| d_{k,1:t})}, \text{ where } \sum_{i=1}^{N} \omega_{k,t}^{i}=1. \label{proportinal_weights}
\end{align} 
We can factorize \eqref{proportinal_weights} such that we obtain the following recursive formula \cite{djuric}
\begin{align}
\omega_{k,t}^{i} \propto \frac{p(d_{k,t} | \vec{a}_{k,t}^{i})p(\vec{a}_{k,t}^{i} | \vec{a}_{k,t-1}^{i})}{q(\vec{a}_{k,t}^{i} | \vec{a}_{k,t-1}^{i},d_{k,t} )}\omega_{k,t-1}^{i}. \label{particleweight} 
\end{align}
In \eqref{particleweight}, we choose the importance function so that the variance of the weights is minimized. By this, we obtain particles that have nonnegligible weights and significantly contribute to \eqref{distribution} \cite{djuric}. In this sense, since $p(\vec{a}_{k,t}^{i}| \vec{a}_{k,t-1}^i)$ provides a small variance for the weights \cite{djuric}, we choose it as our importance function. With this choice, we alter \eqref{particleweight} as follows
\begin{align}
\omega_{k,t}^{i} \propto p(d_{k,t} | \vec{a}_{k,t}^{i})\omega_{k,t-1}^{i}.
\label{particleweight2} 
\end{align}
By \eqref{distribution} and \eqref{particleweight2}, we obtain the state estimate as follows
\begin{align*}
\mathbf{E}[\vec{a}_{k,t}| d_{k,1:t}&] = \int\vec{a}_{k,t}p(\vec{a}_{k,t} | d_{k,1:t}) d\vec{a}_{k,t} \nonumber \\ &\approx \int\vec{a}_{k,t}\sum_{i=1}^{N} \omega_{k,t}^{i}\delta(\vec{a}_{k,t}-\vec{a}_{k,t}^{i}) d\vec{a}_{k,t} 
=\sum_{i=1}^{N} \omega_{k,t}^{i}\vec{a}_{k,t}^{i}.
\end{align*}
Although we choose the importance function to reduce the variance of the weights, the variance inevitably increases over time \cite{djuric}. Hence, we apply the resampling algorithm introduced in \cite{djuric} such that we eliminate the particles with small weights and prevent the variance from increasing.
\subsubsection{Online Training with the DPF Algorithm:\text{}}
In this subsection, we introduce our online training method based on the DPF algorithm when the nodes share information with their neighbors. We employ the Markov Chain Distributed Particle Filter (MCDPF) algorithm \cite{markovconvergence} to train our distributed system. In the MCDPF algorithm, particles move around the network according to the network topology. In every step, each particle can randomly move to another node in the neighborhood of its current node. While randomly moving, the weight of each particle is updated using $p(d_{k,t}|\vec{a}_{k,t})$ at the node $k$, hence, particles use the observations at different nodes. 

\begin{table}
  \centering
  \medskip \medskip
    \caption{Comparison of the computational complexities of the introduced training algorithms for each node $k$. In this table, we also calculate the computational complexity of the SGD based algorithm by deriving exact gradient equations, however, we omit these calculations due to page limit.}
  \resizebox{.6\columnwidth}{.1\columnwidth}{
  \begin{tabular}{| c | c |}
    \hline
    Algorithm & Computational Complexity \\ \hline
    SGD       & $\mathcal{O}(n^{4}+n^{2}p^{2})$ \\ \hline
    DEKF       & $\mathcal{O}(\eta_k( n^8+n^4 p^4))$ \\ \hline
    DPF       & $\mathcal{O}(N(k)(n^2+np))$ \\ \hline
  \end{tabular}
  }
\label{tab:complexity}
\end{table}

Suppose we consider our network as a graph $G=(V,E)$, where the vertices $V$ represent the nodes in our network and the edges $E$ represent the connections between the nodes. In addition to this, we denote the number of visits to each node $k$ in $s$ steps by each particle $i$ as $M^i(k,s)$. Here, each particle moves to one of its neighboring nodes with a certain probability, where the movement
 probabilities of each node to the other nodes are represented by the adjacency matrix, i.e., denoted as $\mathcal{A}$. In this framework, at each visit to each node $k$, each particle multiplies its weight with $p(d_{k,t}|\vec{a}_{k,t})^{\frac{2|E(G)|}{s\eta_{k}}}$ in a run of $s$ steps \cite{markovconvergence}, where $|E(G)|$ is the number of edges in $G$ and $\eta_{k}$ is the degree of the node $k$. From \eqref{particleweight2}, we have the following update for each particle $i$ at the node $k$ after $s$ steps
\begin{align}
w_{k,t}^i =w_{k,t-1}^i \prod_{j=1}^{K} p(d_{j,t} |\vec{a}_{k,t}^i)^{\frac{2|E(G)|}{s\eta_j} M^i(j,s)}. \label{dpf_weight}
\end{align}
We then calculate the posterior distribution at the node $k$ as
\begin{align}
p(\vec{a}_{k,t} | O_{k,t}) \approx \sum_{i=1}^{N} w_{k,t}^i\delta(\vec{a}_{k,t}-\vec{a}_{k,t}^i), \label{dpf_distribution}
\end{align}
where $O_{k,t}$ represents the observations seen by the particles at the node $k$ until $t$ and $w_{k,t}^i $ is obtained from \eqref{dpf_weight}. After we obtain \eqref{dpf_distribution}, we calculate our estimate for $\vec{a}_{k,t}$ as follows
\begin{align}
\mathbf{E}[\vec{a}_{k,t}| O_{k,t}] &= \int\vec{a}_{k,t}p(\vec{a}_{k,t} |O_{k,t}) d\vec{a}_{k,t} \nonumber \\ &\approx \int\vec{a}_{k,t}\sum_{i=1}^{N} \omega_{k,t}^{i}\delta(\vec{a}_{k,t}-\vec{a}_{k,t}^{i}) d\vec{a}_{k,t} \nonumber\\&=\sum_{i=1}^{N} \omega_{k,t}^{i}\vec{a}_{k,t}^{i}.\label{dpf_estimate}
\end{align}
We can obtain the estimate for each node using the same procedure as illustrated in Algorithm \ref{alg2}. In Algorithm \ref{alg2}, $N(j)$ represents the number of particles at the node $j$ and $\mathcal{I}_{i \rightarrow j}$ represents the indices of the particles that move from the node $i$ to the node $j$. Thus, we obtain a distributed training algorithm that guarantees convergence to the optimal centralized parameter estimation as illustrated in Theorem 1. 

\noindent
{\bf Theorem 1: }{\em For each node $k$, let $\vec{a}_{k,t}$ be the bounded state vector with a measurement density function that satisfies the following inequality
\begin{align}
0< p_0\leq p(d_{k,t}|\vec{a}_{k,t}) \leq ||p||_{\infty}< \infty, \label{bound}
\end{align}
where $p_0$ is a constant and
\begin{align*}
 ||p||_{\infty}=\sup_{d_{k,t}}p(d_{k,t}|\vec{a}_{k,t}).
\end{align*}
Then, we have the following convergence results in the MSE sense
\begin{align*}
\sum_{i=1}^{N} \omega_{k,t}^{i}\vec{a}_{k,t}^{i}\rightarrow \mathbf{E}[\vec{a}_{k,t}| \lbrace d_{j,1:t} \rbrace_{j=1}^{K}] \text{ as }  N \rightarrow \infty \text{ and } k \rightarrow \infty  .
\end{align*}}
{\em Proof of Theorem 1. } Using \eqref{bound}, from \cite{markovconvergence}, we obtain
\begin{align}
\mathbf{E}\big[\big(\mathbf{E}[\pi(\vec{a}_{t})|\lbrace d_{j,1:t}& \rbrace_{j=1}^{K}]-\sum_{i=1}^{N} \omega_{k,t}^{i}\pi(\vec{a}_{k,t}^{i})\big)^2 \big] \nonumber \\ &\leq ||\pi||_{\infty}^{2}\bigg(C_t \sqrt{U(s,\upsilon)}+\sqrt{\frac{\varsigma_t}{N}} \bigg)^{2}, \label{theorem1}
\end{align} 
where $\pi$ is a bounded function, $\upsilon$ is the second largest eigenvalue modulus of $\mathcal{A}$, $\varsigma_t$ and $C_t$ are time dependent constants and $U(s,\upsilon)$ is a function of $s$ as described in \cite{markovconvergence} such that $U(s,\upsilon)$ goes to zero as $s$ goes to infinity. Since the state vector $\vec{a}_{k,t}$ is bounded, we can choose $\pi(\vec{a}_{k,t})=\vec{a}_{k,t}$. With this choice, evaluating \eqref{theorem1} as $N$ and $s$ go to infinity yields the results. This concludes our proof. \hfill  $\square$

According to the update procedure illustrated in Algorithm \ref{alg2}, the computational complexity of our training method results in $\mathcal{O}(N(k)(n^2+np))$ computations at each node $k$ due to matrix vector multiplications in \eqref{pfcompact1} and \eqref{pfcompact2} as shown in Table \ref{tab:complexity}. 
\begin{algorithm}
\footnotesize
\caption{Training based on the DPF Algorithm}
\label{alg2}
\begin{algorithmic}[1]
\State{Sample $ \lbrace \vec{a}_{j,t}^i \rbrace_{i=1}^{N(j)} $ from $p(\vec{a}_{t}|\lbrace \vec{a}_{j,t-1}^i \rbrace_{i=1}^{N(j)} ) \text{, } \forall j $}
\State{Set $\lbrace w_{j,t}^i \rbrace_{i=1}^{N(j)} =1 \text{, } \forall j$}
\For{$s$ steps}
\State {Move the particles according to $\mathcal{A}$ }
    \For{$j=1:K$}
       \State{$\lbrace \vec{a}_{j,t}^i \rbrace_{i=1}^{N(j)} \leftarrow\bigcup_{l \in \mathcal{N}_j} \lbrace \vec{a}_{l,t}^i \rbrace_{i \in \mathcal{I}_{l \rightarrow j}} $} 
        \State{$\lbrace w_{j,t}^i \rbrace_{i=1}^{N(j)} \leftarrow\bigcup_{l \in \mathcal{N}_j} \lbrace w_{l,t}^i \rbrace_{i \in \mathcal{I}_{l \rightarrow j}} $} 
        \State{$\lbrace w_{j,t}^i \rbrace_{i=1}^{N(j)} \leftarrow \lbrace w_{j,t}^i \rbrace_{i=1}^{N(j)}p(d_{j,t} |\lbrace \vec{a}_{j,t}^i \rbrace_{i=1}^{N(j)})^{\frac{2|E(G)|}{s\eta_j} }$}
    \EndFor
 \State {\textbf{end for}}
\EndFor
\State {\textbf{end for}}
\For{j=1:K}
\State{Resample $\lbrace \vec{a}_{j,t}^i, w_{j,t}^i \rbrace_{i=1}^{N(j)}$ }
\State{Compute the estimate for node $j$ using \eqref{dpf_estimate}}
\EndFor
\State{\textbf{end for}}
\end{algorithmic}
\end{algorithm}
\section{Simulations}\label{sec:simulations}
We evaluate the performance of the introduced algorithms on different benchmark real datasets. We first consider the prediction performance on Hong Kong exchange rate dataset \cite{hke}. We then evaluate the regression performance on emotion labelled sentence dataset \cite{datasetdigit}. For these experiments, to observe the effects of communication among nodes, we also consider the EKF and PF based algorithms without communication over a network of multiple nodes, where each node trains LSTM based on only its observations. Throughout this section, we denote the EKF and PF based algorithms without communication over a network of multiple nodes as ``EKF" and ``PF", respectively. We also consider the SGD based algorithm without communication over a network of multiple nodes as a benchmark algorithm and denote it by ``SGD".

	 \begin{figure*}[h]
	\centering
	\captionsetup[subfigure]{oneside,margin={1cm,0cm}}
	\begin{subfigure}[t]{0.32\textwidth}
		\centering
		\includegraphics[width=1.15\textwidth, height=1.2\textwidth]{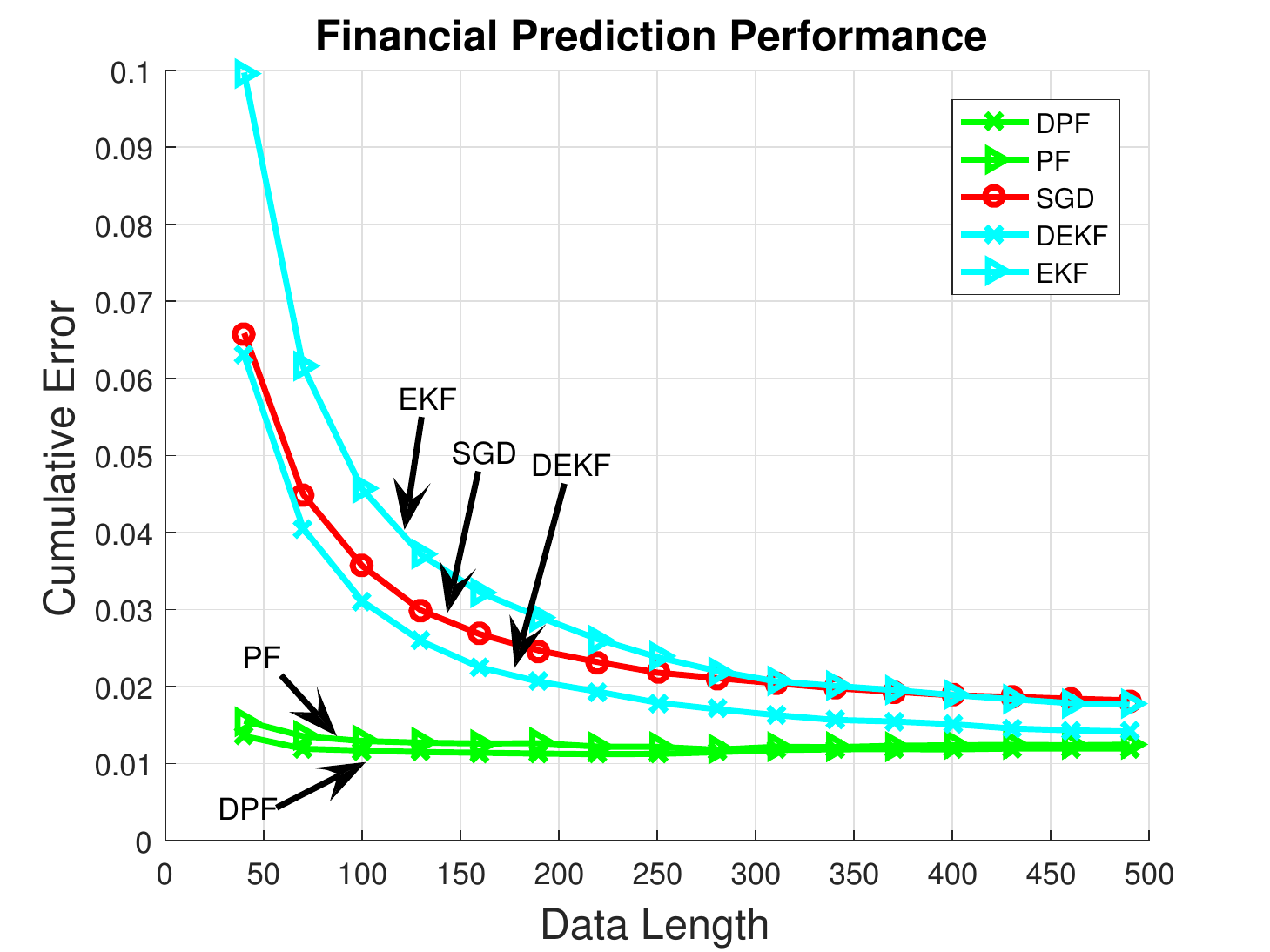}
	\caption{\centering\medskip} \label{hke}
	\end{subfigure}\hspace*{\fill}
		\begin{subfigure}[t]{0.32\textwidth}
		\centering
		\includegraphics[width=1.15\textwidth, height=1.2\textwidth]{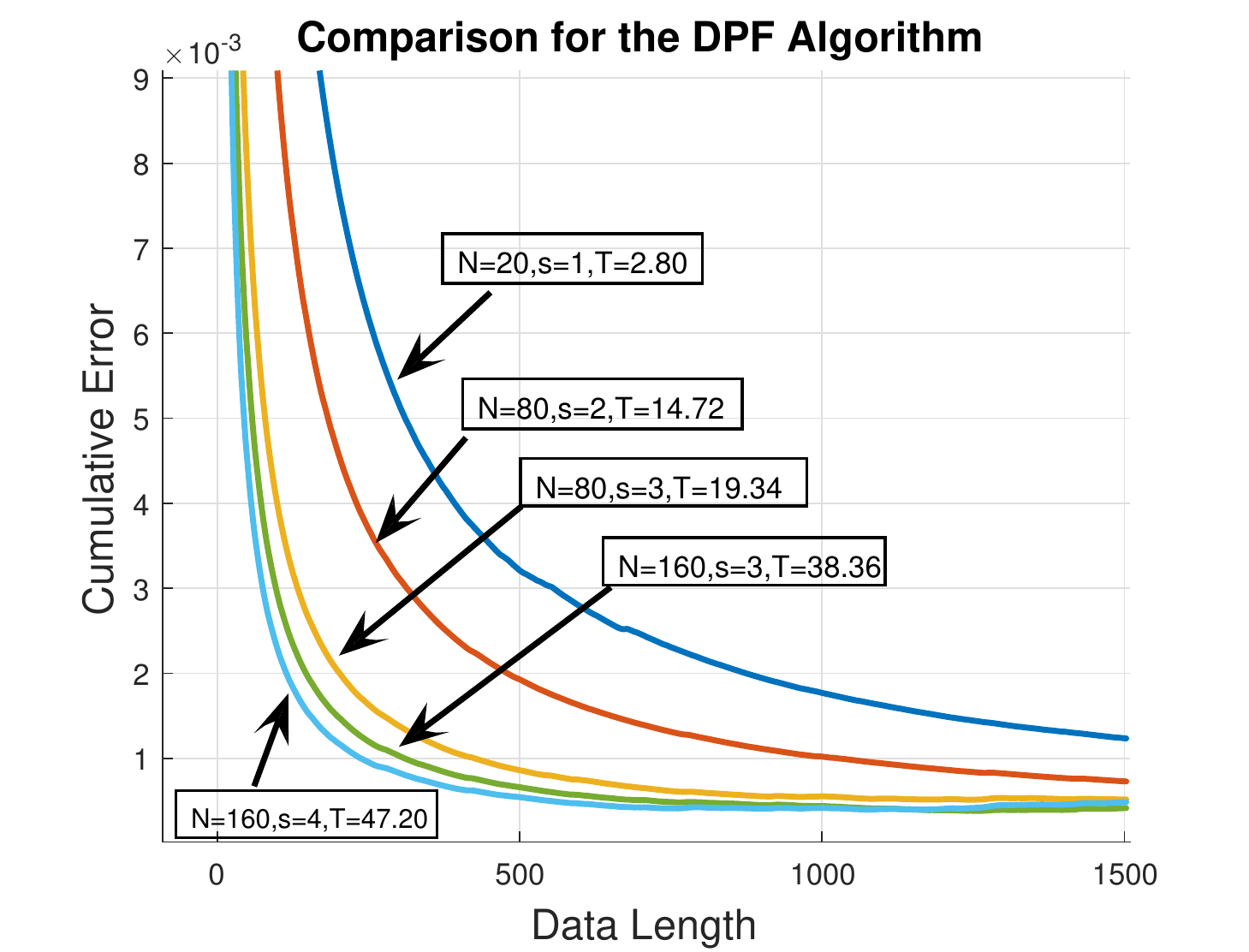}
		\caption{\centering\medskip} \label{varyingkn}
	\end{subfigure} \hspace*{\fill}
		\begin{subfigure}[t]{0.32\textwidth}
		\centering
		\includegraphics[width=1.15\textwidth, height=1.2\textwidth]{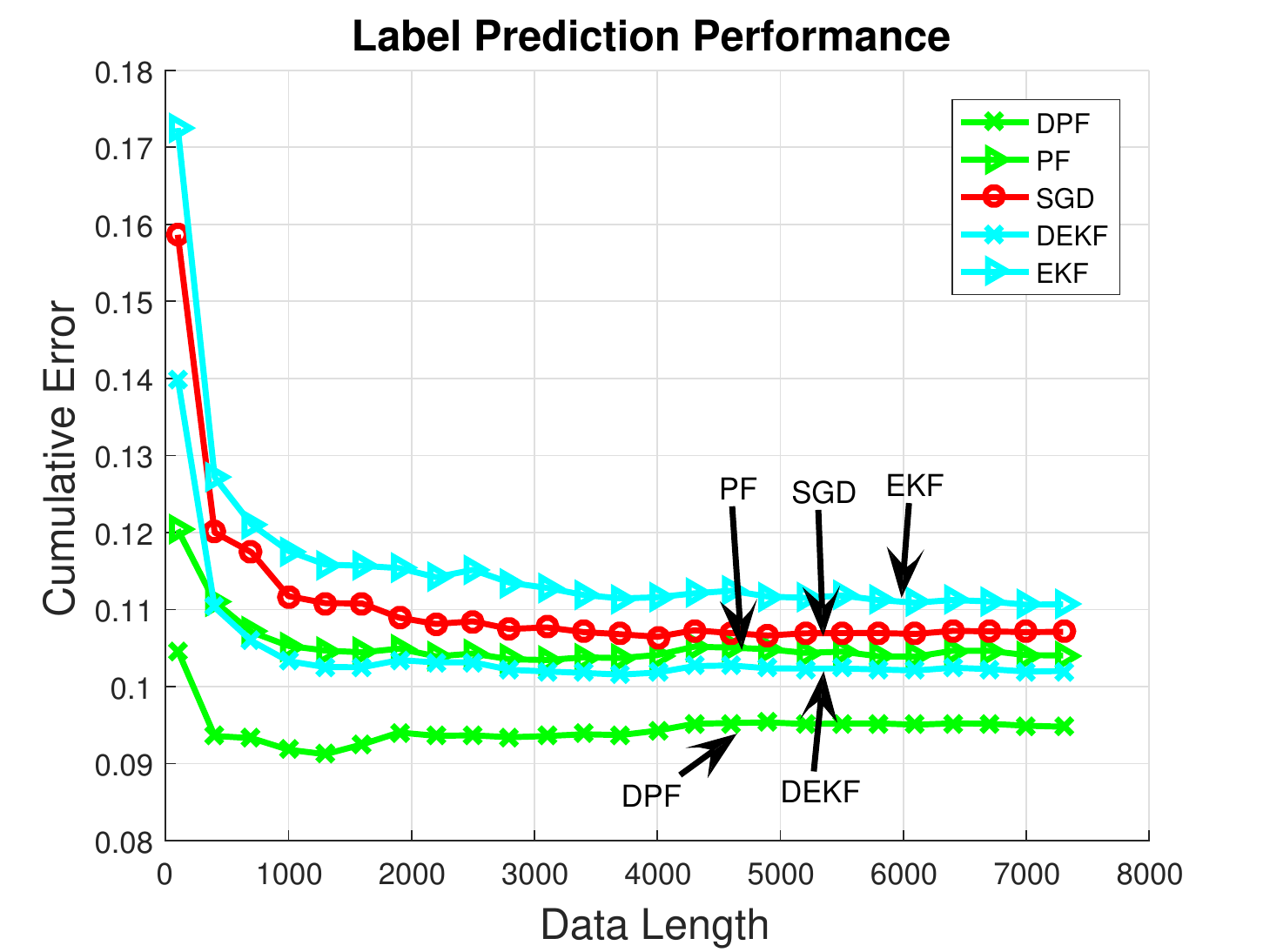}
		\caption{\centering\medskip} \label{digit}
	\end{subfigure} \hspace*{\fill}
	\medskip
	\caption{Error performances (a) over the Hong Kong exchange rate dataset, (b) for different $N$ and $s$ combinations of the DPF based algorithm and (c) over the sentence dataset. In (b), we also provide computation times of the combinations (in seconds), i.e., denoted as $T$, where a computer with i5-6400 processor, 2.7 GHz CPU and 16 GB RAM is used.}\label{figs}
	\end{figure*}
We first consider the Hong Kong exchange rate dataset \cite{hke}. For this dataset, we have the amount of Hong Kong dollars that can buy one United States dollar on certain days. Our aim is to estimate future exchange rate by using the values in the previous two days. In online applications, one can demand a small steady state error or fast convergence rate based on the requirements of application \cite{sayed}. In this experiment, we evaluate the convergence rates of the algorithms. For this purpose, we select the parameters such that the algorithms converge to the same steady state error level. In this setup, we choose the parameters for each node $k$ as follows. Since $\vec{X}_{k,t} \in \mathbb{R}^2$ is our input, we set the output dimension as $n=2$. In addition to this, we consider a network of four nodes. For the PF based algorithms, we choose $N(k)=80$ as the number of particles. Additionally, we select $\vec{\gamma}_{k,t}$ and $\varepsilon_{k,t}$ as zero mean Gaussian random variables with $\mathrm{Cov}[\vec{\gamma}_{k,t}]=0.0004\vec{I}$ and $\mathrm{Var}[\varepsilon_{k,t}]=0.01$, respectively. For the DPF based algorithm, we choose $s=3$ and $\mathcal{A}=[0\text{ }\frac{1}{2}\text{ } 0\text{ } \frac{1}{2}; \frac{1}{2}\text{ } 0 \text{ }\frac{1}{2}\text{ } 0; 0 \text{ }\frac{1}{2} \text{ }0 \text{ }\frac{1}{2}; \frac{1}{2}\text{ } 0 \text{ }\frac{1}{2} \text{ }0]$.
For the EKF based algorithms, we select $\vec{\Sigma}_{k,0|0}=0.0004 \vec{I}$, $\vec{Q}_{k,t}=0.0004 \vec{I}$ and $R_{k,t}=0.01$. Moreover, according to \eqref{eq:phi}, the weights between nodes are calculated as $1/3$. For the SGD based algorithm, we set the learning rate as $\mu=0.1$. In Fig. \ref{hke}, we illustrate the prediction performance of the algorithms. Due to the highly nonlinear structure of our model, the EKF and DEKF based algorithms have slower convergence compared to the other algorithms. Moreover, due to only exploiting the first order gradient information, the SGD based algorithm has also slower convergence compared to the PF based algorithms. Unlike the SGD and EKF based methods, the PF based algorithms perform parameter estimation through a high performance gradient free density estimation technique, hence, they converge much faster to the final MSE level. Among the PF based methods, due to its distributed structure the DPF based algorithm has the fastest convergence rate.

In order to demonstrate the effects of the number of particles $N$ and the number of Markov steps $s$, we perform another experiment using the Hong Kong exchange rate dataset. In this experiment, we use the same setting with the previous case except $\mathrm{Cov}[\vec{\gamma}_{k,t}]=0.0001\vec{I}$, $\vec{\Sigma}_{k,0|0}=0.0001 \vec{I}$ and $\vec{Q}_{k,t}=0.0001 \vec{I}$. In Fig. \ref{varyingkn}, we observe that as $s$ and $N$ increase, the DPF based algorithm obtains a faster convergence rate and a lower final MSE value. However, as $s$ and $N$ increase, the marginal performance improvement becomes smaller with respect to the previous $s$ and $N$ values. Furthermore, the computation time of the algorithm increases with increasing $s$ and $N$ values. Thus, after a certain selection, a further increase does not worth the additional computational load. Therefore, we use $N(k)=80$ and $s=3$ in our previous simulation.

Other than the Hong Kong exchange rate dataset, we consider the emotion labelled sentence dataset \cite{datasetdigit}. In this dataset, we have the vector representation of each word in an emotion labelled sentence. In this experiment, we evaluate the steady state error performance of the algorithms. Thus, we choose the parameters such that the convergence rate of the algorithms are similar. To provide this setup, we select the parameters for each node $k$ as follows. Since the number of words varies from sentence to sentence in this case, we have a variable length input regressor, i.e., defined as $\vec{X}_{k,t} \in \mathbb{R}^{2 \times m_t}$, where $m_t$ represents the number of words in a sentence. For the other parameters, we use the same setting with the Hong Kong exchange rate dataset except $N(k)=50$, $\mathrm{Cov}[\vec{\gamma}_{k,t}]=(0.025)^2\vec{I}$, $\vec{\Sigma}_{k,0|0}=(0.025)^2 \vec{I}$, $\vec{Q}_{k,t}=(0.025)^2 \vec{I}$ and $\mu =0.055$. In Fig. \ref{digit}, we illustrate the label prediction performance of the algorithms. Again due to the highly nonlinear structure of our model, the EKF based algorithm has the highest steady state error value. Additionally, the SGD based algorithm also has a high final MSE value compared to the other algorithms. Furthermore, the DEKF based algorithm achieves a lower final MSE value than the PF based method thanks to its distributed structure. However, since the DPF based method utilizes a powerful gradient free density estimation method while effectively sharing information between the neighboring nodes, it achieves a much smaller steady state error value.
\section{Concluding Remarks}\label{sec:conclusion}
We studied online training of the LSTM architecture in a distributed network of nodes for regression and introduced online distributed training algorithms for variable length data sequences. We first proposed a generic LSTM based model for variable length data inputs. In order to train this model, we put the model equations in a nonlinear state space form. Based on this form, we introduced distributed extended Kalman and particle filtering based online training algorithms. In this way, we obtain effective training algorithms for our LSTM based model. Here, our distributed particle filtering algorithm guarantees convergence to the optimal centralized parameter estimation in the MSE sense under certain conditions. We achieve this performance with communication and computational complexity in the order of the first order methods \cite{tsoi}. Through simulations involving real life and financial data, we illustrate significant performance improvements with respect to the state of the art methods \cite{ekf_lstm2,rtrl}.

\begin{spacing}{.87}
\small
\bibliographystyle{IEEEtran}
\bibliography{my_ref}
\end{spacing}
\end{document}